\documentclass[twoside,11pt]{8ssmmp} 

\usepackage{amsfonts}
\usepackage{amssymb}
\usepackage{amsmath}
\usepackage{fancyhdr}    
\usepackage{equations}
\usepackage{epic}
\usepackage{eepic}
\usepackage{epsf}
\usepackage{graphicx}
\usepackage{rotating}
\usepackage{hyperref}

\newcommand\rf[1]{(\ref{eq:#1})}
\newcommand\lab[1]{\label{eq:#1}}
\newcommand\nonu{\nonumber}
\newcommand\br{\begin{eqnarray}}
\newcommand\er{\end{eqnarray}}
\newcommand\be{\begin{equation}}
\newcommand\ee{\end{equation}}

\newcommand\lb{\lbrack}
\newcommand\rb{\rbrack}

\renewcommand\({\left(}
\renewcommand\){\right)}

\newcommand\bc{\begin{center}}
\newcommand\ec{\end{center}}

\newcommand\partder[2]{\frac{{\partial {#1}}}{{\partial {#2}}}}

\renewcommand\a{\alpha}

\newcommand\eps{\epsilon}
\newcommand\vareps{\varepsilon}

\newcommand\G{\Gamma}

\newcommand\h{\frac{1}{2}}
\renewcommand\k{\kappa}
\renewcommand\l{\lambda}

\newcommand\m{\mu}
\newcommand\n{\nu}

\newcommand\vp{\varphi}
\renewcommand\P{\Phi}
\newcommand\pa{\partial}

\newcommand\wti{\widetilde}

\newcommand\cH{{\mathcal H}}

\newcommand{\ct}[1]{\cite{#1}}
\newcommand{\bib}[1]{\bibitem{#1}}

\newcommand\PRL[3]{\textsl{Phys. Rev. Lett.} \textbf{#1} (#2) #3}

\newcommand\PRD[3]{\textsl{Phys. Rev.} \textbf{D#1} (#2) #3}

\newcommand\PLB[3]{\textsl{Phys. Lett.} \textbf{#1B} (#2) #3}
\newcommand\CQG[3]{\textsl{Class. Quantum Grav.} \textbf{#1} (#2) #3}

\newcommand\IJMPD[3]{\textsl{Int. J. Mod. Phys.} \textbf{D#1} (#2) #3}

\newcommand\MPLA[3]{\textsl{Mod. Phys. Lett.} \textbf{A#1} (#2) #3}

\newcommand\udot{\stackrel{.}{u}}

\newcommand\Adot{\stackrel{.}{A}}
\newcommand\Bdot{\stackrel{.}{B}}
\newcommand\Hdot{\stackrel{.}{H}}
\renewcommand{\headrulewidth}{0.4pt}

\begin{document}
 \baselineskip=11pt

\title{Unification of Inflation and Dark Energy from Spontaneous Breaking of Scale 
Invariance
}
\author{\bf{Eduardo Guendelman}\hspace{.25mm}\thanks{\,e-mail address:
guendel@bgu.ac.il}
\\ \normalsize{Physics Department, Ben Gurion University of the Negev}\\
\normalsize{Beer Sheva, Israel} \vspace{2mm} \\ 
\bf{Emil Nissimov and Svetlana Pacheva}\hspace{.25mm}\thanks{\,e-mail
address: nissimov@inrne.bas.bg, svetlana@inrne.bas.bg}
\\ \normalsize{Institute for Nuclear Research and Nuclear Energy}\\
\normalsize{Bulgarian Academy of Sciences, Sofia, Bulgaria} }

\date{}

\maketitle

\begin{abstract}
We propose a new class of gravity-matter models defined in terms of two
independent non-Riemannian volume forms (alternative generally covariant
integration measure densities) on the spacetime manifold. For the matter we choose
appropriate scalar field potentials of exponential form so that the full
gravity-matter system is invariant under global Weyl-scale symmetry. Solution
of the pertinent equations of motion produce two dimensionful integration
constants which spontaneously break global Weyl-scale invariance. In the resulting
effective Einstein-frame gravity-matter system we obtain an effective
potential for the scalar matter field which has an interesting cosmological
application, namely, it allows for a unified description of both an early
universe inflation and present day dark energy. 
\end{abstract}

\section{Introduction}
\label{intro}

A component of the energy-momentum tensor of matter which is proportional to 
the spacetime metric tensor, with the proportionality constant being indeed 
exactly or approximately a spacetime constant, has been widely discussed and its 
consequences understood, but the possible origin of such terms remains a subject 
of hot discussion.

One can transfer such energy-momentum tensor component to the left hand side of 
Einstein's equations and then it can be considered as belonging to the gravity part.
This was the way indeed how Einstein introduced such contribution and named it the
``cosmological constant term''.

More recently it has been invoked as a fundamental component of the energy density 
of both the early universe and of the present universe. Nowadays we call such 
component ``vacuum energy density''. The vacuum energy density has been used as 
the source of a possible inflationary phase of the early 
universe (the pioneering papers on the subject are \ct{early-univ-1};
for a non-technical review and a good collection of further references on 
different aspects of inflation see Ref.\ct{early-univ-2}; for a more technical
review see Ref.\ct{early-univ-3}). Inflation provides an attractive scenario for
solving some of the fundamental puzzles of the standard Big Bang model, like 
the horizon and the flatness problems (third ref.\ct{early-univ-1}) as well as 
providing a framework for sensible calculations of primordial density 
perturbations (for a review, see the book \ct{dodelson}).

Also, with the discovery of the accelerating expansion of the present 
universe (for reviews of this subject, see for example \ct{accel-exp,accel-exp-2})
it appears plausible that a small vacuum energy density, 
usually referred in this case as ``dark energy'', is also present even today. 
Because of this discovery the cosmological constant problem (CCP) has evolved
from the ``Old Cosmological Constant Problem'' \ct{weinberg-1}, where physicists 
were concerned with explaining why the observed vacuum energy density of the 
universe nowadays is vanishing, to a different type of CCP -- the 
``New Cosmological Constant Problem'' \cite{weinberg-2}. Namely, the problem now is  
to explain why the vacuum energy density of the current universe is very small 
rather than being zero. 

These two vacuum energy densities, the one of inflation and the other of the 
universe nowadays, have however a totally different scale. One then wonders 
how cosmological evolution may naturally interpolate between such two 
apparently quite distinctive physical situations.

The possibility of continuously connecting an inflationary phase to a slowly 
accelerating universe through the evolution of a single scalar field -- the
{\em quintessential inflation scenario} -- has been first studied in 
Ref.\ct{peebles-vilenkin}. Also, carefully constructed F(R) models can yield 
both an early time inflationary epoch and a late time de Sitter phase with 
vastly different values of effective vacuum energies \ct{starobinsky-2}.
For a recent proposal of a quintessential
inflation mechanism based on ``variable gravity'' model \ct{wetterich} and for
extensive list of references to earlier work on quintessential inflation,
see Ref.\ct{murzakulov-etal}.

In the present letter we propose a new theoretical framework where the
quintessential inflation scenario is explicitly realized in a natural way.

The main idea of our current approach comes from 
Refs.\ct{TMT-orig-1,TMT-orig-2,TMT-orig-3} 
(for recent developments, see Refs.\ct{TMT-recent}), where some of us
have proposed a new class of gravity-matter theories based on the idea that the 
action integral may contain a new metric-independent integration measure
density, \textsl{i.e.}, an alternative non-Riemannian volume form on the
spacetime manifold defined in terms of an auxiliary antisymmetric gauge
field of maximal rank. The latter formalism yields various new interesting results
in all types of known generally coordinate-invariant theories:

\begin{itemize}
\item
(i) $D=4$-dimensional models of gravity and matter fields containing 
the new measure of integration appear to be promising candidates for resolution 
of the dark energy and dark matter problems, the fifth force problem, 
and a natural mechanism for spontaneous breakdown of global Weyl-scale symmetry
\ct{TMT-orig-1}-\ct{TMT-recent}.
\item
(ii) Study of reparametrization invariant theories of extended objects 
(strings and branes) based on employing of a modified non-Riemannian 
world-sheet/world-volume integration measure \ct{mstring} leads to dynamically 
induced variable string/brane tension and to string models of non-abelian 
confinement.
\item
(iii) Study in Refs.\ct{susy-break} of modified supergravity models with an
alternative non-Riemannian volume form on the spacetime manifold produces some
outstanding new features: 
(a) This new formalism applied to minimal $N=1$ supergravity 
naturally triggers the appearance of a dynamically generated cosmological constant
as an arbitrary integration constant, which signifies a new explicit
mechanism of spontaneous (dynamical) breaking of supersymmetry;
(b) Applying the same formalism to anti-de Sitter supergravity allows us to 
appropriately choose the above mentioned arbitrary 
integration constant so as to obtain simultaneously a very small effective
observable cosmological constant as well as a very large physical gravitino mass.
\end{itemize}

We now extend the above formalism employing two (instead of only one) modified 
non-Riemannian volume-forms on the underlying spacetime to construct new type of
gravity-matter models producing interesting cosmological implications
relating inflationary and slowly accelerating phases of the universe.

\section{Gravity-Matter Models With Two Independent \\ Non-Riemannian Volume-Forms}
\label{TMMT}

We shall consider the following non-standard gravity-matter system with an action 
of the general form (for simplicity we will use units where the Newton
constant is taken as $G_N = 1/16\pi$):
\be
S = \int d^4 x\,\P_1 (A) \Bigl\lb R + L^{(1)} \Bigr\rb +  
\int d^4 x\,\P_2 (B) \Bigl\lb L^{(2)} + \frac{\P (H)}{\sqrt{-g}}\Bigr\rb \; ,
\lab{TMMT}
\ee
with the following notations:

\begin{itemize}
\item
$\P_{1}(A)$ and $\P_2 (B)$ are two independent non-Riemannian volume-forms, 
\textsl{i.e.}, generally covariant integration measure densities on the underlying
spacetime manifold:
\be
\P_1 (A) = \frac{1}{3!}\vareps^{\m\n\k\l} \pa_\m A_{\n\k\l} \quad ,\quad
\P_2 (B) = \frac{1}{3!}\vareps^{\m\n\k\l} \pa_\m B_{\n\k\l} \; ,
\lab{Phi-1-2}
\ee
defined in terms of field-strengths of two auxiliary 3-index antisymmetric
tensor gauge fields. $\P_{1,2}$ take over the role of the standard
Riemannian integration measure density 
$\sqrt{-g} \equiv \sqrt{-\det\Vert g_{\m\n}\Vert}$ in terms of the spacetime
metric $g_{\m\n}$.
\item
$R = g^{\m\n} R_{\m\n}(\G)$ and $R_{\m\n}(\G)$ are the scalar curvature and the 
Ricci tensor in the first-order (Palatini) formalism, where the affine
connection $\G^\m_{\n\l}$ is \textsl{a priori} independent of the metric $g_{\m\n}$.
\item
$L^{(1,2)}$ denote two different Lagrangians with matter fields, to be
specified below.
\item
$\P (H)$ indicate the dual field strength of a third auxiliary 3-index antisymmetric
tensor gauge field:
\be
\P (H) = \frac{1}{3!}\vareps^{\m\n\k\l} \pa_\m H_{\n\k\l}; ,
\lab{Phi-H}
\ee
whose presence is crucial for non-triviality of the model. 
\end{itemize}

For the matter Lagrangians we take the scalar field ones:
\be
L^{(1)} = -\h g^{\m\n} \pa_\m \vp \pa_\n \vp - V(\vp) \quad ,\quad
L^{(2)} = U(\vp) \;\; ({\rm no ~kinetic ~term}) \; .
\lab{L-1-2}
\ee
We now observe that the original action \rf{TMMT} is invariant under
global Weyl-scale transformations:
\br
g_{\m\n} \to \l g_{\m\n} \;\; ,\;\; \vp \to \vp - \frac{1}{\a}\ln \l \;\;,
\nonu \\
A_{\m\n\k} \to \l A_{\m\n\k} \;\; ,\;\; B_{\m\n\k} \to \l^2 B_{\m\n\k}
\;\; ,\;\; H_{\m\n\k} \to H_{\m\n\k} \; ,
\lab{scale-transf}
\er
where $\a$ is a dimensionful positive parameter, provided we choose the scalar 
field potentials in \rf{L-1-2} in the form (similar to the choice \ct{TMT-orig-1}):
\be
V(\vp) = f_1 \exp \{-\a\vp\} \quad ,\quad U(\vp) = f_2 \exp \{-2\a\vp\} \; .
\lab{V-U-eqs}
\ee

Variation of \rf{TMMT} w.r.t. $\G^\m_{\n\l}$ gives (following the derivation
in \ct{TMT-orig-1}):
\be
\G^\m_{\n\l} = \G^\m_{\n\l}({\bar g}) = 
\h {\bar g}^{\m\k}\(\pa_\n {\bar g}_{\l\k} + \pa_\l {\bar g}_{\n\k} 
- \pa_\k {\bar g}_{\n\l}\) \; ,
\lab{G-eq}
\ee
where ${\bar g}_{\m\n}$ is the Weyl-rescaled metric:
\be
{\bar g}_{\m\n} = \chi_1 g_{\m\n} \;\; ,\;\; 
\chi_1 \equiv \frac{\P_1 (A)}{\sqrt{-g}} \; .
\lab{bar-g}
\ee

Variation of the action \rf{TMMT} w.r.t. auxiliary tensor gauge fields
$A_{\m\n\l}$, $B_{\m\n\l}$ and $H_{\m\n\l}$ yields the equations:
\be
\pa_\m \Bigl\lb R + L^{(1)} \Bigr\rb = 0 \quad, \quad
\pa_\m \Bigl\lb L^{(2)} + \frac{\P (H)}{\sqrt{-g}}\Bigr\rb = 0 \quad, \quad
\pa_\m \Bigl(\frac{\P_2 (B)}{\sqrt{-g}}\Bigr) = 0 \; ,
\lab{A-B-H-eqs}
\ee
whose solutions read:
\br
\frac{\P_2 (B)}{\sqrt{-g}} = \chi_2 = {\rm const} \quad ,\quad
R + L^{(1)} = - M_1 = {\rm const} \; ,
\nonu \\
L^{(2)} + \frac{\P (H)}{\sqrt{-g}} = - M_2  = {\rm const} \; .
\lab{integr-const}
\er
Here $M_1$ and $M_2$ are arbitrary dimensionful and $\chi_2$
arbitrary dimensionless integration constants. 
The appearance of $M_1,\, M_2$ signifies {\em dynamical spontaneous
breakdown} of global Weyl-scale invariance under \rf{scale-transf} due to the scale
non-invariant solutions (second and third ones) in \rf{integr-const}.

Varying \rf{TMMT} w.r.t. $g_{\m\n}$ and using relations \rf{integr-const} 
we have:
\be
\chi_1 \Bigl\lb R_{\m\n} + \partder{}{g^{\m\n}} L^{(1)}\Bigr\rb -
\h \chi_2 \Bigl\lb T^{(2)}_{\m\n} + g_{\m\n}M_2\Bigr\rb = 0 \; ,
\lab{pre-einstein-eqs}
\ee
where $\chi_1$ and $\chi_2$ are defined in \rf{bar-g} and first relation 
\rf{integr-const}, and $T^{(2)}_{\m\n}$ is the energy-momentum tensor of the second
matter Lagrangian with the standard definitions:
\be
T^{(1,2)}_{\m\n} = g_{\m\n} L^{(1,2)} - 2 \partder{}{g^{\m\n}} L^{(1,2)} \; .
\lab{EM-tensor}
\ee
Using second relation \rf{integr-const} and \rf{EM-tensor}, 
Eqs.\rf{pre-einstein-eqs} can be put in the form:
\be
R_{\m\n} - \h g_{\m\n}R = \h \Bigl\lb T^{(1)}_{\m\n} + g_{\m\n} M_1
+ \frac{\chi_2}{\chi_1}\( T^{(2)}_{\m\n} + g_{\m\n} M_2\)\Bigr\rb \; .
\lab{einstein-like-eqs}
\ee
Taking the trace of Eqs.\rf{einstein-like-eqs} and using again second relation 
\rf{integr-const} we solve for the scale factor $\chi_1$:
\be
\chi_1 = 2 \chi_2 \frac{U(\vp)+M_2}{V(\vp) - M_1} \; .
\lab{chi-1}
\ee
Now, taking into account \rf{bar-g} and \rf{chi-1} we can bring
Eqs.\rf{einstein-like-eqs} into the standard form of Einstein equations for
the metric ${\bar g}_{\m\n}$ \rf{bar-g} , \textsl{i.e.}, the Einstein frame
equations: 
\be
R_{\m\n}({\bar g}) - \h {\bar g}_{\m\n} R({\bar g}) = \h T^{\rm eff}_{\m\n}
\lab{eff=einstein-eqs}
\ee
with energy-momentum tensor corresponding (according to \rf{EM-tensor}) to
the following effective scalar field Lagrangian:
\be
L_{\rm eff} = - \h {\bar g}^{\m\n} \pa_\m \vp \pa_\n \vp - U_{\rm eff} (\vp) \; ,
\lab{L-eff}
\ee
where the effective scalar field potential reads:
\be
U_{\rm eff} (\vp) = \frac{\bigl(V(\vp)-M_1\bigr)^2}{4\chi_2 \bigl(U(\vp)+M_2\bigr)}
= \frac{\(f_1 e^{-\a\vp}-M_1\)^2}{4\chi_2\,\(f_2 e^{-2\a\vp}+M_2\)} \; .
\lab{U-eff}
\ee
\section{Canonical Hamiltonian Treatment}
\label{hamiltonian}

Before proceeding to the cosmological implications of the new gravity-matter
model based on two non-Riemannian spacetime volume forms let us briefly
discuss the application of the canonical Hamiltonian formalism to \rf{TMMT},
which will elucidate the proper physical meaning of the arbitrary
integration constants $\chi_2,\, M_1,\, M_2$ \rf{integr-const} encountered in the
previous section.

For convenience let us introduce the following short-hand notations for the
field-strengths \rf{Phi-1-2}, \rf{Phi-H} of the auxiliary 3-index antisymmetric gauge 
fields $A_{\m\n\l},\, B_{\m\n\l},\, H_{\m\n\l}$ (the dot indicating time-derivative): 
\br
\P_1 (A) = \Adot + \pa_i A^i \quad, \quad 
A = \frac{1}{3!} \vareps^{ijk} A_{ijk} \;\; ,\;\;
A^i = - \h \vareps^{ijk} A_{0jk} \; ,
\lab{A-can} \\
\P_1 (B) = \Bdot + \pa_i B^i \quad, \quad 
B = \frac{1}{3!} \vareps^{ijk} B_{ijk} \;\; ,\;\;
B^i = - \h \vareps^{ijk} B_{0jk} \; ,
\lab{B-can} \\
\P (H) = \Hdot + \pa_i H^i \quad, \quad 
H = \frac{1}{3!} \vareps^{ijk} H_{ijk} \;\; ,\;\;
H^i = - \h \vareps^{ijk} H_{0jk} \; ,
\lab{H-can}
\er
For the pertinent canonical momenta we have:
\br
\pi_A = R + L^{(1)} \equiv {\wti L}_1 (u,\udot) \;\; ,\;\;
\pi_B = L^{(2)} (u,\udot) + \frac{1}{\sqrt{-g}}(\Hdot + \pa_i H^i) \; ,
\nonu \\
\pi_H = \frac{1}{\sqrt{-g}}(\Bdot + \pa_i B^i) \; ,
\lab{can-momenta-aux}
\er
where $(u,\udot)$ collectively denote the set of the basic gravity-matter
canonical variables ($(u)=(g_{\m\n}, \vp, \ldots)$) and their velocities, 
and:
\be
\pi_{A^i} = 0 \quad,\quad \pi_{B^i} = 0 \quad,\quad \pi_{H^i} = 0 \; ,
\lab{can-momenta-zero}
\ee
the latter implying that $A^i, B^i, H^i$ will in fact appear as Lagrange multipliers
for certain first-class Hamiltonian constraints 
(see Eqs.\rf{pi-A-const}-\rf{pi-B-pi-H-const} below). For the canonical momenta
conjugated to the basic gravity-matter canonical variables we have 
(using last relation \rf{can-momenta-aux}):
\be
p_u = (\Adot + \pa_i A^i) \frac{\pa}{\pa \udot} {\wti L}_1 (u,\udot) + 
\pi_H \sqrt{-g} \frac{\pa}{\pa \udot} L^{(2)} (u,\udot) \; .
\lab{can-momenta-u}
\ee

Now, from relations \rf{can-momenta-aux}, \rf{can-momenta-u} we obtain the velocities 
$\udot,\Adot,\Bdot,\Hdot$ as functions of canonically conjugate momenta 
$\udot = \udot (u,\pi_A,\pi_B,\pi_H)$ \textsl{etc.} Taking into account
\rf{can-momenta-aux}-\rf{can-momenta-zero} (and the short-hand notations
\rf{A-can}-\rf{H-can}) the canonical Hamiltonian corresponding to \rf{TMMT}:
\br
\cH = p_u \udot + \pa_A \Adot + \pa_B \Bdot + \pa_H \Hdot -
(\Adot + \pa_i A^i) {\wti L}_1 (u,\udot) 
\nonu \\
- \pi_H \sqrt{-g} \Bigl\lb L^{(2)}(u,\udot) + 
\frac{1}{\sqrt{-g}}(\Hdot + \pa_i H^i) \Bigr\rb
\lab{can-hamiltonian}
\er
acquires the following form as function of the canonically conjugated variables
(here $\udot = \udot (u,\pi_A,\pi_B,\pi_H)$):
\br
\cH = p_u \udot - \pi_H \sqrt{-g} L^{(2)}(u,\udot) + \sqrt{-g} \pi_H \pi_B
\nonu \\
- \pa_i A^i \pi_A - \pa_i B^i \pi_B - \pa_i H^i \pi_H \; .
\lab{can-hamiltonian-final}
\er
We are interested only in the canonical Hamiltonian structure related to the
auxiliary antisymmetric tensor gauge fields.
From the second line in \rf{can-hamiltonian-final} we deduce that indeed 
$A^i, B^i, H^i$ are Lagrange multipliers for the first-class Hamiltonian constraints:
\be
\pa_i \pi_A = 0 \;\; \to\;\; \pi_A = - M_1 = {\rm const} \; ,
\lab{pi-A-const}
\ee
and similarly:
\be
\pi_B = - M_2 = {\rm const} \quad ,\quad \pi_H = \chi_2 = {\rm const} \; ,
\lab{pi-B-pi-H-const}
\ee
which are the canonical Hamiltonian counterparts of Lagrangian constraint
equations of motion \rf{integr-const}.

Thus, the canonical Hamiltonian treatment of \rf{TMMT} reveals the meaning
of the auxiliary 3-index antisymmetric tensor gauge fields
$A_{\m\n\l},\, B_{\m\n\l},\, H_{\m\n\l}$ -- building blocks of
the non-Riemannian spacetime volume-form formulation of the modified gravity-matter
model \rf{TMMT}. Namely, the canonical momenta $\pi_A,\, \pi_B,\, \pi_H$ 
conjugated to the ``magnetic'' parts $A,B,H$ \rf{A-can}-\rf{H-can}
of the auxiliary 3-index antisymmetric tensor gauge fields are constrained
through Dirac first-class constraints \rf{pi-A-const}-\rf{pi-B-pi-H-const}
to be constants identified with the arbitrary 
integration constants $\chi_2,\, M_1,\, M_2$ \rf{integr-const} arising within the 
Lagrangian formulation of the model. The canonical momenta 
$\pi_A^i,\, \pi_B^i,\, \pi_H^i$ conjugated to the ``electric'' parts $A^i,B^i,H^i$ 
\rf{A-can}-\rf{H-can} of the auxiliary 3-index antisymmetric tensor gauge field
are vanishing \rf{can-momenta-zero} which makes the latter canonical Lagrange 
multipliers for the above Dirac first-class constraints.

\section{Implications for Cosmology}
\label{cosmolog}

The effective scalar potential \rf{U-eff} possesses the following remarkable
property.
For large negative and large positive values of $\vp$ $U_{\rm eff} (\vp)$ 
exponentially fast approaches two infinitely large flat regions (which we
will denote as $(\mp)$ flat regions, respectively) with smooth transition
between them:
\br
U_{\rm eff} (\vp) \to \frac{f_1^2}{4\chi_2 f_2} \quad {\rm for}\;\;
\vp \to -\infty \; ,
\nonu \\
U_{\rm eff} (\vp) \to \frac{M_1^2}{4\chi_2 M_2} \quad {\rm for}\;\;
\vp \to +\infty \; .
\lab{flat-regions}
\er
The shape of $U_{\rm eff} (\vp)$ depicted on Fig.1. 

\begin{figure}
\begin{center}
\includegraphics{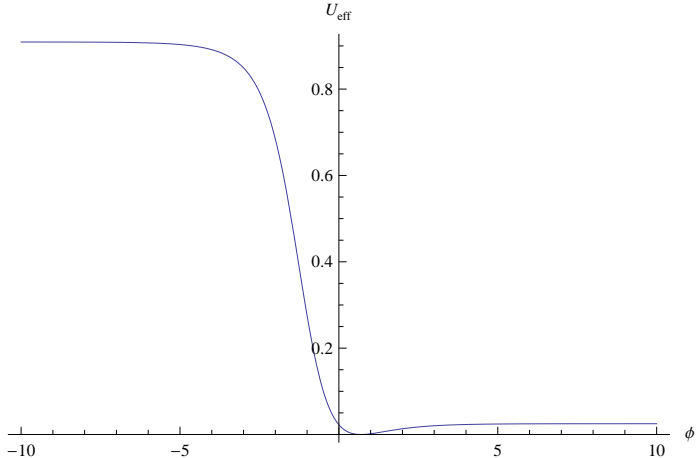}
\caption{Shape of the effective scalar potential $U_{\rm eff} (\vp)$ \rf{U-eff}.}
\end{center}
\end{figure}

In the original gravity-matter models with only one 
non-Riemannian volume form \ct{TMT-orig-1} one obtains upon spontaneous breakdown
of global Weyl-scale symmetry only one flat region of the effective scalar potential, so
that this simple model does not meet the requirement for unification of
inflation and dark energy.

Let us point out that in the context of the original
modified-measure gravity-matter theories (with only one non-Riemannian
integration measure density) it is possible to obtain two flat regions by
means of adding an $\epsilon R^2$ term as shown in \ct{TMT-orig-3}. This is,
however, achieved at the price of creating a non-canonical kinetic term for the
scalar field which substantially complicates the theory and its particle content 
interpretation (see remarks on this point below).

In the present case we derived an effective scalar potential $U_{\rm eff} (\vp)$ 
\rf{U-eff} with {\em two infinitely large flat regions} while the kinetic term of 
the scalar field remained canonical. In the course of the derivation we obtained 
three integration constants $\chi_2,\, M_1,\, M_2$ \rf{integr-const}, two of them 
($M_1,\, M_2$) triggering spontaneous breakdown of the original global Weyl-scale 
symmetry \rf{scale-transf}. 
These integration constants can be appropriately adjusted so as to get the shape 
of the effective scalar potential as depicted on Fig.1.

The cosmological picture suggested by Fig.1 is evident. The universe starts
from a large negative value of $\vp$, then slow rolls the $(-)$ flat region to the
left whose height: 
\be
U_{\rm eff} (\vp) \simeq U_{(-)} = \frac{f_1^2}{4\chi_2 f_2} 
\lab{U-minus}
\ee
upon appropriate choice of $f_1,\, f_2$ can be made very large corresponding
to the vacuum energy density in the inflationary phase. After that there
is an abrupt fall to $U_{\rm eff} = 0$ where particle creation is obtained
from rapidly varying $\vp (t)$. The scalar field comes down with very high
kinetic energy in the region of $U_{\rm eff} \simeq 0$, certainly higher
than the value of $U_{\rm eff}$ in the $(+)$ flat region to the right:
\be
U_{\rm eff} (\vp) \simeq U_{(+)} = \frac{M_1^2}{4\chi_2 M_2} \; , 
\lab{U-plus}
\ee
which upon appropriate choice of the scales of $M_1,\, M_2$ (see below) can be 
made to correspond to the correct value of the current vacuum energy density. 
So $\vp (t)$ ``climbs'' the latter low barrier and continues to evolve in the
$\vp \to +\infty$ direction. Thus, on the $(+)$ flat region we have a slow
rolling scalar which produces approximately the dark energy equation of state
($\rho \simeq - p$, with very small $\rho \simeq U_{(+)} = \frac{M_1^2}{4\chi_2 M_2}$)
explaining the present day dark energy phase.

Indeed, taking the integration constant $\chi_2 \sim 1$, and choosing the
scales of the scalar potential \rf{U-eff} coupling constants 
$M_1 \sim M^4_{EW}$ and $M_2 \sim M^4_{Pl}$, where $M_{EW},\, M_{Pl}$ are
the electroweak and Plank scales, respectively, we are then naturally led to
a very small vacuum energy density of the order:
\be
\rho \simeq U_{(+)} \sim M^8_{EW}/M^4_{Pl} \sim 10^{-120} M^4_{Pl} \; ,
\lab{U-plus-magnitude}
\ee
which  gives the right order of magnitude for the present epoche's vacuum energy 
density as already recognized in Ref.\ct{arkani-hamed}.

In a parallel work \ct{emergent} we have generalized the model \rf{TMMT} by
including a gravitational $R^2$ term so as to preserve the original global
Weyl-scale symmetry \rf{scale-transf}:
\be
S = \int d^4 x\,\P_1 (A) \Bigl\lb R + L^{(1)} \Bigr\rb +  
\int d^4 x\,\P_2 (B) \Bigl\lb L^{(2)} + \eps R^2 + 
\frac{\P (H)}{\sqrt{-g}}\Bigr\rb \; .
\lab{TMMT-2}
\ee
The analysis of the model \rf{TMMT-2} goes along similar lines as described in
Sections 2 and 3 above, where in addition we find for a definite parameter range 
a non-singular ``emergent universe'' solution which describes an initial phase
of universe's evolution that precedes the inflationary phase. It was also
realized in \ct{emergent} that upon taking the order of magnitude for the
coupling constants in the effective scalar potential 
$f_1 \sim f_2 \sim (10^{-2} M_{Pl})^4$, then the order of magnitude of the
vacuum energy density of the early universe $U_{(-)}$ \rf{U-minus} becomes:
\be
U_{(-)} \sim f_1^2/f_2 \sim 10^{-8} M_{Pl}^4 \; ,
\lab{U-minus-magnitude}
\ee
which conforms to the BICEP2 experiment \ct{bicep2} and Planck Collaboration
data \ct{Planck} implying the energy scale of inflation of order 
$10^{-2} M_{Pl}$. Nevertheless, as shown in \ct{emergent},  the result for the
tensor-to-scalar ratio $r$ obtained within the model \rf{TMMT-2} conforms 
to the data of the Planck Collaboration \ct{Planck} rather than 
BICEP2 \ct{bicep2}.

\section*{Acknowledgments.} 
E.G. and E.N. are sincerely grateful to Prof. Branko Dragovich and the organizers
of the {\em Eight Meeting in Modern Mathematical Physics} in Belgrade for cordial 
hospitality. We are thankful to Alexander Kaganovich for stimulating discussions,
as well as to Ramon Herrera and Pedro Labrana for a fruitful collaboration on the
extension of the current project.
We gratefully acknowledge support of our collaboration through the academic exchange 
agreement between the Ben-Gurion University and the Bulgarian Academy of Sciences.
E.N. and S.P. are partially supported by Bulgarian NSF Grant \textsl{DFNI T02/6}.
S.P. has received partial support from European COST action MP-1210.


\end{document}